\documentclass[final,times,twocolumn]{elsarticle}
 \biboptions{comma,sort&compress}
\usepackage{here}
 \usepackage{graphicx}
  \usepackage{epsfig}

\def\nin{\noindent}
\def\beq{\begin{equation}}
\def\eeq{\end{equation}}
\def\bea{\begin{eqnarray}}
\def\eea{\end{eqnarray}}

\def\la{\langle}
\def\ra{\rangle}

\journal{Nuc. Phys. (Proc. Suppl.)}

\begin{document}

\begin{frontmatter}



\title{$\tau^-\to\eta\pi^-\pi^0\nu_\tau$ and $\sigma(e^+e^-\to \eta\pi^+\pi^-)$ at low energies}

 \author[label1]{Pablo Roig\corref{cor1}}
  \address[label1]{Laboratoire
de Physique Th\'eorique\\
Universit\'e de Paris-Sud XI, B\^atiment 210,
 91405. Orsay cedex, France.}
\cortext[cor1]{Speaker}
\ead{pablo.roig@th.u-psud.fr}



\begin{abstract}

\noindent
We analyze the hadronization structure of $\tau \to \eta \pi^- \pi^0 \nu_\tau$ decays. In the isospin limit only the vector current contributes to this 
process. We compute the relevant form factor within Resonance Chiral Theory, at leading order in the $1/N_C$ expansion, and considering only the
contribution of the lightest vector resonances. The couplings in the resonance theory are constrained by imposing the asymptotic behaviour of 
vector spectral functions ruled by $QCD$. We reproduce the branching ratio of this mode and predict the low-energy behaviour of 
$\sigma\left(e^+e^-\to\eta\pi^+\pi^-\right)$ using isospin symmetry.

\end{abstract}




\end{frontmatter}


\section{Introduction} \label{Intro}
\nin
Hadronic decays of the $\tau$ are a clean scenario were to learn features about the hadronization of QCD currents in its non-perturbative regime \cite{Tau} and 
the decays involving the $\eta$ meson are particularly interesting because of the selection rules that apply \cite{Pich:1987qq} in the isospin limit. Within this 
approximation, it is seen that the $\eta$ and the vector current, $\cal{V}_\mu$, have opposed G-parity than the $\pi$ and the axial-vector current, $\cal{A_\mu}$. Then, the decay 
$\tau\to\eta\pi^-\nu_\tau$ can only be produced in the SM as an isospin violating effect, since it has opposite G-parity to the participating vector 
current. The processes that we study here, $\tau \to \eta(p_1) \pi^-(p_2) \pi^0(p_3) \nu_\tau$, can be considered as due only to $\cal{V}_\mu$ to a very good 
approximation.\\
The decay amplitude for these decays may be written as:
\begin{equation} \label{Mgraltau}
\mathcal{M}\,=\,-\frac{G_F}{\sqrt{2}}\,V_{\mathrm{ud}}\,\overline{u}_{\nu_\tau}\gamma^\mu(1-\gamma_5)\,u_\tau \mathcal{H}_\mu\,,
\end{equation}
where we will consider the hadron tensor given by the $\cal{V}_\mu$ contribution only:
\begin{equation} \label{Hmugral}
\mathcal{H}_\mu = \la \left\lbrace  P(p_i)\right\rbrace_{i=1}^3 |\mathcal{V}_\mu e^{i\mathcal{L}_{QCD}}|0\ra\,,
\end{equation}
with
\begin{eqnarray} \label{Hmu3m}
 \mathcal{H}_\mu = i \,\varepsilon_{\mu\nu\varrho\sigma}\,p_1^\nu\, p_2^\varrho\, p_3^{\sigma}\,F(Q^2,\,s_1,\,s_2)\,,\nonumber\\
 Q_\mu \, = \, (p_1\,+\,p_2\,+\,p_3)_\mu ,\, s_i = (Q\,-\,p_i)^2\,.
\end{eqnarray}

\section{Theoretical framework} \label{Theory}
\nin Since QCD is non-perturbative for $Q^2\lesssim M_\tau^2$, it is not possible to go further in a model independent way. However, it would be desirable 
to keep as many QCD properties as possible. $M_\tau$ is large enough to prevent the application of $\chi PT$ \cite{ChPT} for all the spectra \cite{Colangelo:1996hs}, 
however in the low-energy limit one should recover its results. The expansion parameter in $\chi PT$ ceases to be valid at $E\sim M_\rho$ and the inverse of 
the number of colours in QCD is a useful alternative to build the expansion upon \cite{Nc} and helps formulate a Lagrangian theory, $R\chi T$, where the 
resonances that mediate hadronic $\tau$ decays become active degrees of freedom \cite{RChT} and the known short-distance QCD behaviour \cite{BrodskyLepage} 
has been demanded to the Green functions \cite{VVP, GFs} and associated form factors.\\
For any phenomenological study it will be essential to provide the resonances with a proper energy-dependence width, that we obtain within $R\chi T$\cite{widths}. 
This program has been applied successfully to explain the phenomenology of many two- and three-meson $\tau$ decay channels~\cite{PhenoTau, KKpi}.\\
In the case at hand, $\tau^{-} \to \eta \pi^{-} \pi^0 \nu_\tau$, only the vector current contributes in the isospin limit. Our formalism includes the Wess-Zumino 
term \cite{WZW} as in $\chi PT$. In addition, one considers the one-resonance mediated diagrams with an odd-parity term: the couplings of a vector and a pseudoscalar 
to the vector source \cite{VVP} ($c_i$ couplings) and those in which a vector resonance couples to three mesons \cite{Roig:2007yp} ($g_i$ couplings). 
Two-resonance mediated diagrams are also accounted for in this negative parity sector \cite{VVP} (the coefficients of the $VVP$ operators are called $d_i$). 
13 of the 18 Lagrangian couplings appear in the process under study.\\
Demanding the known ultraviolet behaviour of the imaginary part of the vector-vector correlator \cite{Floratos:1978jb} one obtains relations among the couplings 
\footnote{Noteworthy, they are consistent with all phenomenological studies mentioned before.} that allow the theory to keep some predictive power. In this case, only 4 couplings remain unknown after this step. We read the value of the free combination 
of $g_i$ couplings from Ref.~\cite{KKpi} and rely on the value obtained in Ref.~\cite{VVP} of a $d_i$ whose coefficient vanishes in the chiral limit ($d_3$). 
Therefore, only 2 coefficients are still free.\\
\section{Associated Phenomenology} \label{Pheno}
\begin{figure}[h!]
 \begin{center}
\includegraphics[scale=0.42]{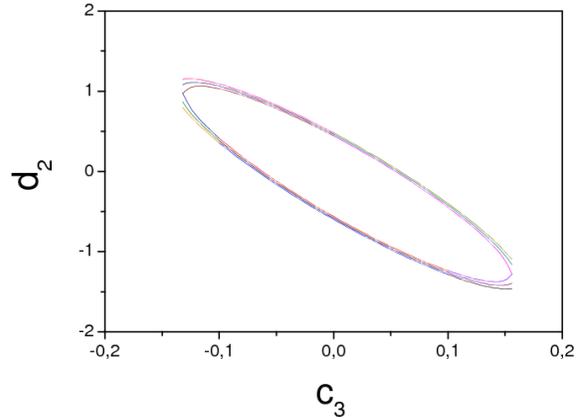}
\caption[]{\label{Fig1} \small{One-sigma contours for the branching ratio of the decay $\tau^{-} \rightarrow \eta \pi^{-} \pi^{0} \nu_{\tau}$ in the 
$c_3$-$d_2$ plane.}}
 \end{center}
\end{figure}

\begin{figure}[h!]
 \begin{center}
\includegraphics[scale=0.33, angle=-90]{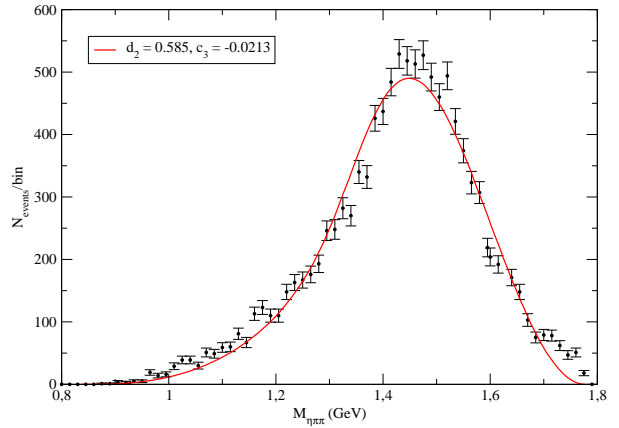}
\caption[]{\label{Fig2} \small{Fit to Belle data \cite{Inami:2008ar} on $\tau^{-} \rightarrow \eta \pi^{-} \pi^{0} \nu_{\tau}$.}}
 \end{center}
\end{figure}

\begin{figure}[h!]
 \begin{center}
  \vspace*{0.5cm}
\includegraphics[scale=0.33, angle=-90]{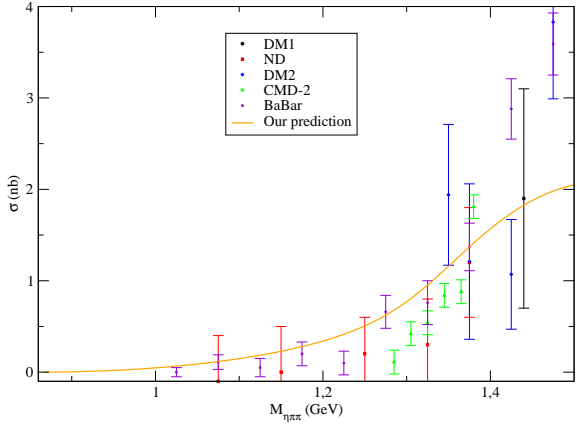}
\caption[]{\label{Fig3} \small{Prediction for the low-energy behaviour of $\sigma(e^+e^-\to \eta\pi^+\pi^-)$ based on the $\tau^{-} \rightarrow \eta \pi^{-} \pi^{0} \nu_{\tau}$ 
decays analysis compared to  DM1 \cite{Delcourt:1982sj}, ND \cite{Druzhinin:1986dk}, DM2 \cite{Antonelli:1988fw}, CMD-2 \cite{Akhmetshin:2000wv} and 
BaBar \cite{Aubert:2007ef} data.}}
 \end{center}
\end{figure}

\nin We reproduce \cite{Paper} the PDG branching ratio \cite{PDG}, $(1.39\pm0.10)\times10^{−3}$, for natural values of both couplings ($c_3$ and $d_2$) as it is 
displayed in Fig.~\ref{Fig1}.\\
In addition to this, we have fit these two couplings to Belle data \cite{Inami:2008ar} on the number of events per bin versus the invariant mass of the hadron 
system as it is represented in Fig \ref{Fig2}. We have checked that the best fit is obtained with the value of $d_2$ assumed in Sect. \ref{Theory}. Moreover, we have also verified that the inclusion 
of a $\rho'$ does not improve sensibly the fit and does not produce the bump structure that is observed near the endpoint.\\ \\
Finally, using isospin symmetry, it is possible to relate the low-energy $\sigma(e^+e^-\to \eta\pi^+\pi^-)$ to $\Gamma(\tau^{-} \rightarrow \eta \pi^{-} \pi^{0} \nu_{\tau})$ 
\cite{Paper}. Then, our analysis for $\tau^{-} \to \eta \pi^{-} \pi^0 \nu_\tau$ decays allows to predict the low-energy limit of this cross-section, as displayed 
in Fig.~\ref{Fig3}. \footnote{One can proceed conversely and use the data on $e^+e^-$ annihilation into hadrons to predict the corresponding semileptonic 
tau decays \cite{Eidelman:1990pb, Cherepanov:2009zz}.}\\ \\
\section{Conclusions} \label{Conclusions}
\nin We have studied the decay $\tau^{-} \to \eta \pi^{-} \pi^0 \nu_\tau$ within the framework of Resonance Chiral Theory guided by the large-$N_C$ expansion 
of $QCD$, the low-energy limit given by $\chi PT$ and the appropriate asymptotic behaviour of the vector form factor that helps to fix most of the initially 
unknown couplings. Indeed only two remain free after completing this procedure and having used information acquired in the previous related analyses. 
We reproduce the PDG branching ratio for this mode and study its spectral function assisted by Belle data.\\
\hspace*{0.5cm} These are flourishing days for this branch of Physics, where the data samples collected at CLEO, BaBar and Belle through the years have 
allowed very precise studies and current and forthcoming experimental results from them, BES-III, VEPP and hopefully super-B and super-$\tau$-charm factories 
will demand a dedicated effort both on the theory description and on the Monte Carlo event generation algorithms \cite{Actis:2010gg}. These hadron matrix 
elements will be implemented \cite{Olga} in the Monte Carlo Generator for $\tau$ decays TAUOLA \cite{TAUOLA}. Using isospin symmetry, we provide a prediction 
for the low-energy behaviour of $\sigma(e^+e^-\to \eta\pi^+\pi^-)$ that may be of interest for the hadronic matrix element in the PHOKHARA \cite{PHOKHARA} 
Monte Carlo generator.\\

\section*{Acknowledgements}
\nin
I congratulate Stephan Narison and all his team on the enjoyable QCD2010 Conference where I have benefited from discussions on this topic with Z.~H.~Guo and 
P.~Masjuan. The work reported here has been done in collaboration with D.~G\'omez Dumm and A.~Pich. 
We thank H.~Czyz and S.~Eidelman for their interest in our work and useful correspondence. I am indebted to K.~Hayasaka and K.~Inami for giving me access to Belle data for 
our analysis. I also would like to thank D.~G.~Dumm for his comments on the manuscript. I acknowledge the financial support of a Marie Curie ESR Contract (FLAVIAnet). 
This work has been supported in part by the EU MRTN-CT-2006-035482 (FLAVIAnet), by MEC (Spain) under grant FPA2007-60323, by the Spanish Consolider-Ingenio 
2010 Programme CPAN (CSD2007-00042).






\end{document}